\documentclass[sigconf]{acmart}
\AtBeginDocument{%
  \providecommand\BibTeX{{%
    \normalfont B\kern-0.5em{\scshape i\kern-0.25em b}\kern-0.8em\TeX}}}

\setcopyright{acmcopyright}
\copyrightyear{2023}
\acmYear{2023}
\setcopyright{acmlicensed}\acmConference[WWW '23]{Proceedings of the ACM Web Conference 2023}{May 1--5, 2023}{Austin, TX, USA}
\acmBooktitle{Proceedings of the ACM Web Conference 2023 (WWW '23), May 1--5, 2023, Austin, TX, USA}
\acmPrice{15.00}
\acmDOI{10.1145/3543507.3583218}
\acmISBN{978-1-4503-9416-1/23/04}

\usepackage{subcaption}
\usepackage{multirow}
\usepackage{float}
\usepackage{url}
\usepackage{dblfloatfix}
\usepackage{footnote}
\usepackage{subcaption}

\DeclareCaptionLabelFormat{r-parens}{Figure 2{#2:}} 

\begin{document}

\title{Longitudinal Assessment of Reference Quality on Wikipedia}

\author{Aitolkyn Baigutanova}
\authornote{Both authors led this work together.}
\orcid{0000-0002-9852-4157}
\affiliation{
  \institution{KAIST}
  \city{Daejeon}
  \country{South Korea}
}
\email{aitolkyn.b@kaist.ac.kr}

\author{Jaehyeon Myung}
\authornotemark[1]
\orcid{0000-0003-2294-9762}
\affiliation{%
  \institution{KAIST}
  \city{Daejeon}
  \country{South Korea}
}
\email{mjhbest@kaist.ac.kr}

\author{Diego Saez-Trumper}
\orcid{0000-0002-7679-5423}
\affiliation{%
  \institution{Wikimedia Foundation}
  \city{Barcelona}
  \country{Spain}
}
\email{diego@wikimedia.org}

\author{Ai-Jou Chou}
\orcid{0000-0002-5269-7167}
\affiliation{%
  \institution{Wikimedia Foundation}
  \city{London}
  \country{United Kingdom}
}
\email{qwanqwanro@gmail.com}

\author{Miriam Redi}
\orcid{0000-0002-0581-0251}
\affiliation{
  \institution{Wikimedia Foundation}
  \city{London}
  \country{United Kingdom}
}
\email{mredi@wikimedia.org}

\author{Changwook Jung}
\orcid{0000-0002-5618-5230}
\affiliation{%
  \institution{KAIST}
  \city{Daejeon}
  \country{South Korea}
}
\email{changwook.jung@kaist.ac.kr}

\author{Meeyoung Cha}
\orcid{0000-0003-4085-9648}
\affiliation{%
 \institution{IBS \& KAIST}
 \city{Daejeon}
 \country{South Korea}
 }
 \email{mcha@ibs.re.kr}


\renewcommand{\shortauthors}{Baigutanova, et al.}




\begin{abstract}
Wikipedia plays a crucial role in the integrity of the Web. This work analyzes the reliability of this global encyclopedia through the lens of its references. We operationalize the notion of reference quality by defining \textbf{reference need (RN)}, i.e., the percentage of sentences missing a citation, and \textbf{reference risk (RR)}, i.e., the proportion of non-authoritative references. We release Citation Detective, a tool for automatically calculating the RN score, and discover that the RN score has dropped by 20 percent point in the last decade, with more than half of verifiable statements now accompanying references. The RR score has remained below 1\% over the years as a result of the efforts of the community to eliminate unreliable references. We propose pairing novice and experienced editors on the same Wikipedia article as a strategy to enhance reference quality. Our quasi-experiment indicates that such a co-editing experience can result in a lasting advantage in identifying unreliable sources in future edits. As Wikipedia is frequently used as the ground truth for numerous Web applications, our findings and suggestions on its reliability can have a far-reaching impact. We discuss the possibility of other Web services adopting Wiki-style user collaboration to eliminate unreliable content.
\end{abstract}

\maketitle

\section{Introduction}

Wikipedia is a freely available online encyclopedia maintained by the Web community's collaborative effort. Although anyone can freely edit content, Wikipedia is generally seen to provide ``high-quality'' information as it employs several important guiding content policies. One of them is verifiability~\cite{verifiability}, which requires contributors to support their edits with appropriate references to ensure that other people using the encyclopedia can check where the information came from via inline citations~\cite{piccardi2020quantifying}. Given the sheer volume of information published on the website that is reused by major search engines and social media, the question of verifiability has become more critical~\cite{redi2019citation}. Unfortunately, no existing work has ever investigated the extent to which the encyclopedia preserves the integrity of information on the Web.

We present a methodology to measure Wikipedia's content verifiability through its reference quality by defining two metrics of reference quality. The first is the \textit{reference need (RN)} index, which represents the proportion of citation-missing sentences among those that require a citation. The second is the \textit{reference risk (RR)}, which measures the proportion of non-authoritative sources according to the classification in the perennial source list~\cite{PerennialSource}, a community-driven label for references. These two indicators help assess the status quo of reference quality on Wikipedia over a decade from 2010 to 2020. Along with the paper, we also release \textit{Citation Detective}, a public tool for computing reference need on Wikipedia. This tool helps obtain the ratio of sentences that need a citation for any page and the historical revisions of a page. This tool builds on our previous work~\cite{redi2019citation}, which proposed the initial machine-learning classifier and its technical validation.\footnote{For data and code, refer to: \url{https://github.com/aitolkyn99/wiki_reference_quality}}

Our large-scale analysis using the two reference quality metrics indicates the longitudinal trends of Wikipedia's reliability over the past decade. Temporal analysis shows how the ratios of unreliable sources and citation-missing sentences have decreased over time. The trend differed by topical categories of articles; Wikipedia pages about biology and the earth, for example, had better reference quality than pages about people's lives and the military. The reference quality of the most popular articles was considerably greater than that of a random page, both in terms of RN and RR scores. Our findings on the community efforts around maintaining the list of risky sources are discussed in light of other community-level collaborations that try to eliminate such references.

We further seek a strategy to improve reference quality by examining the interactions between expert and novice editors. A quasi-experiment using propensity score matching (PSM) that pairs revisions from comparable editing environments (i.e., revision byte size and reference score) suggests the importance of prior editing experience. Editors with more experience tend to make better changes in terms of reference 
quality by adding missing references and removing potentially risky sources. Furthermore, our PSM analysis shows that new editors who have co-edited an article with an experienced editor on the same day are more likely to avoid risky references in future edits compared to their peers who have not.

Our findings on the longitudinal assessment of reference quality, as well as the strategy to facilitate co-editing between editors with varying levels of experience, have implications for Web service design. Wikipedia content is contributed collaboratively rather than individually, and our analysis demonstrates the value of this approach in terms of content verifiability. Other systems may consider empowering users, as in the Wikipedia community, thereby fostering positive interaction among users and aiding them in identifying unreliable information. As more services transition toward a collaborative framework, the insights shared in this paper have ramifications for the Web.

\section{Related Work}

References on Wikipedia have been studied directly and indirectly from various perspectives. \smallskip

\noindent
\textbf{Reliability of the Web:}
While most content reliability studies focus on social networks~\cite{pierri2019false}, just a few studies assess the trustworthiness of Websites~\cite{meel2020fake,web1,web2}. Given the role above of Wikipedia as one of the main hubs for the Web~\cite{piccardi2021value}, the online encyclopedia plays an important role in finding trustable content on the Internet~\cite{saez2019online}. 
\smallskip

\noindent
\textbf{Impact of references on Wikipedia}: Being a tertiary source \cite{WS}, Wikipedia's quality is affected mainly by the references used in its articles. Reliable references have become central to Wikipedia as it acts as an educational resource in under-resourced environments~\cite{lemmerich2019world}. Not only are Wikipedia readers affected by the quality of content, but a large set of automated fact-checking datasets and solutions rely on it as ground-truth~\cite{chernyavskiy2021whatthewikifact,karagiannis2019mining,sathe2020automated,thorne2018fever,trokhymovych2021wikicheck}, giving the online encyclopedia a central role in the AI ecosystem. 

Like web services, outbound traffic from clicking a referenced link may take up a small portion of Wikipedia. However, containing references has important implications. Wikipedia acts as a hub of the Internet, driving traffic to other websites and having a measurable economic impact~\cite{piccardi2021value}. For instance, past work has shown a correlation between the chance of being cited on Wikipedia and the impact factor of publication venues~\cite{kousha2017wikipedia,teplitskiy2017amplifying}. \smallskip

\noindent
\textbf{Reference quality:} Several studies have examined the reliability of Wikipedia references in the past. For example, recent research has investigated the reliance of Wikipedia on scientific journal sources~\cite{puyu2021mapofscience} and political polarization of referenced news sources~\cite{puyu2022polarization}. Several researchers examined the role of topic and language in source reliability~\cite{lewoniewski2022}. Regarding reliability and editors, it has been shown that editors who frequently edit an article play a critical role in improving the verifiability of the articles~\cite{referencedynamics}. A systematic review synthesizing a diverse body of work on Wikipedia's content reliability noted a positive trend in the overall quality of citations~\cite{mesgari2015sum}.

There are datasets containing different characterizations of references, such as the presence of identifiers like \emph{DOIs} and \emph{ISBNs}~\cite{singh2021wikipedia,halfaker2018citations}. While most research has focused on English Wikipedia, some studies have examined the cross-lingual presence of DOIs to observe similarities and differences in references used across languages~\cite{lewoniewski2017analysis}. A dataset of all the references existent in English Wikipedia up to June 2019 was recently published and described the references, their history, and the contributing editors~\cite{zagovora2020updated}. Our work goes one step further, studying how the editor's past editing history impacts the quantity and quality of the references they cite.  
\smallskip

\noindent
\textbf{Fact-checking and claim verification:} Since any user can edit Wikipedia content, argument mining and verification become essential. Studies have shown that editors primarily add references to fortify the information they edit and prevent their edits from being removed~\cite{forte2018}. Numerous studies have released datasets for claim verification in specific domains, such as scientific claims~\cite{Wadden2020FactOF, Wadden2022SciFactOpenTO}. Studies like~\cite{Daxenberger2017WhatIT} target cross-domain verification and show shared characteristics among these datasets. Our work focuses on broad topics to account for the wide variety of topics covered by Wikipedia articles. Compared to other work, we also consider the guidelines employed by Wikipedia's verifiability policy.

Our work is based on the Citation Need model~\cite{redi2019citation}, which constructs the Wikipedia verifiability taxonomy and uses machine learning to determine whether a sentence needs a citation. We expand the model into an end-to-end inference pipeline that is usable by the Wikipedia community and examine the trends seen in a decade's worth of data. In addition to extending previous work, we use a crowd-sourced list of domains compiled by Wikipedia editors who manually evaluated the credibility of those sources to assess the reliability~\cite{PerennialSource}.

\section{Data}

Our analysis is based on a large dataset of articles, citations, references, and their labels collected from the English Wikipedia.

\subsection{Wikipedia's Editing Process }

Articles on Wikipedia have unique identifiers, and every edit is logged as a \textit{revision} along with a timestamp, the editing user's information, and a comment describing the edit, as shown in the Appendix. Edits made by users without the login credential are marked as \textit{anonymous} revisions. When bots make edits, this information is stored. If edits involve only minor changes to an article (e.g., fixing a typo), it is labeled as a \textit{minor} revision and, otherwise, a \textit{major} revision. The creation process and the editing history of every Wikipedia article are publicly available.

\subsection{Topical Categories}

We augmented the data for each edit by classifying articles by topics following the methodology in~\cite{johnson2021language} that lists 64 topical categories of Wikipedia articles, as defined in the ORES taxonomy~\cite{ORES}, a topic routing based on machine learning. The adopted ORES topic routing is a hierarchical classification starting with four meta-categories: Culture, History and Society, Geography, and STEM. We used the second level of topical categories. For example, for topics such as Culture.Media.Books and STEM.Biology, the second level is Media and Biology, respectively.

\subsection{Labeling References by Reliability}

References in Wikipedia are core to the verifiability of the content. Therefore, they are expected to be from reliable published sources. However, enforcing this policy is not straightforward, and neither is measuring it. The Wikipedia editing community has been classifying the reliability of the sources that have been frequently questioned. This classification is known as the \textit{perennial sources list} \cite{PerennialSource} and includes 356 web domains as of September 2021.

\begin{figure*}[t!]
 \centering \includegraphics[width=0.7\linewidth]{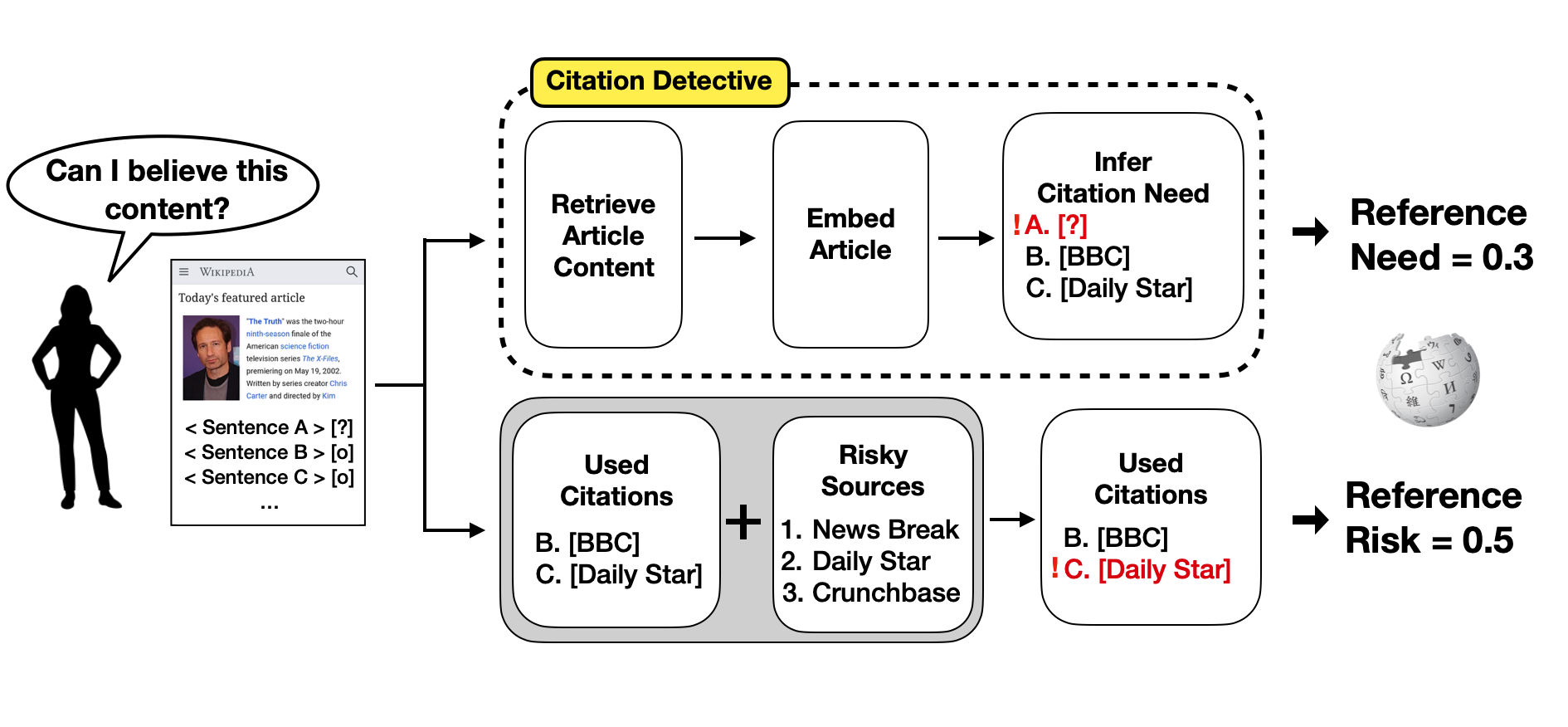}
    \Description[Wikipedia's Reference Quality scores computation]{An example of the pipeline to compute two reference quality scores: Reference Need (RN) and Reference Risk (RR). The first image in the pipeline shows a sample article with three sentences [A, B, and C], where B and C have corresponding references and A is missing a reference. Next, the pipeline is divided into two ways for calculating RN (top-side) and RR (bottom-side). In the top-side box, the Citation Detective tool identifies the need for citations in the three sentences and outputs the ratio of missing references in the sample article. In this example, while all sentences need citations, only sentences B and C have them, so the RN score is 0.3. In the bottom-side box, existing citations [B-BBC and C-Daily Star] are checked for the occurrence of risky sources. Since Daily Star is classified as a risky source in the list, the RN score is 0.5.}
    \vspace*{-7mm}
    \caption{Illustration of how the two reference quality scores are calculated. Reference Need (the proportion of sentences missing citations) is calculated by processing the article content with our \textit{Citation Detective} tool. Reference Risk (the proportion of non-authoritative citations) is computed by extracting all citations used in the article and checking for risky sources.
}
\label{fig:idea}
\end{figure*}

The list is non-exhaustive, as only those sources that have been the subject of repeated community discussion are verified. It consists of five label types: generally reliable, non-consensus, generally unreliable, deprecated, and blacklisted sources. The first three categories are considered context-dependent, and their credibility depends on the area of knowledge in which they are used (e.g., \texttt{arXiv.com} is listed in the ``generally unreliable'' category). However, the latter two categories (i.e., blacklisted and deprecated) should not be used in any context. Among the 356 web domain sources in the list, only 18 are blacklisted, and 36 are deprecated. Examples of perennial sources for each label are provided in the Appendix. This current research utilizes two bottom categories, as they are suggested to be prohibited in general:
\begin{itemize}
    \item \textbf{Deprecated:} A community consensus has been reached to deprecate these sources. All sources in this category are considered generally unreliable; hence, their use is prohibited. Despite this, these sources sometimes appear in discussions of controversial self-descriptions, although reliable secondary sources are still preferred. Examples include \texttt{last.fm}, \texttt{baike.baidu.com}, and \texttt{crunchbase.com}.

    \item \textbf{Blacklisted:} These sources are on the spam-related blacklist due to persistent abuse, usually in the form of embedded external links. Examples include \texttt{thepointsguy.com}, \texttt{zoominfo.com}, and \texttt{naturalnews.com}.
\end{itemize}

\subsection{Datasets}

We built four datasets from the English edition of Wikipedia. All our datasets are released for replication.
\begin{itemize}
\item \textbf{Current} dataset includes the latest versions of 6,476,790 webpages that existed on English Wikipedia as of March 2022. 

\item \textbf{Random} revision dataset includes 3,177,963 revisions of randomly sampled 20K pages.

\item \textbf{Top} revision dataset includes 23,802,067 revisions of 10K pages that received the highest number of total page views in the English Wikipedia within the analyzed period, as computed by Wikimedia's Pageviews API \cite{WikiPageview}.
\end{itemize}

Every editing revision is logged with the following metadata: revision id, timestamp, user id, prior revision count of the editing user, user type anonymous or not, bot or not, page id,  revision byte size difference compared to the prior revision, and revision type minor or not. Using the public Wikipedia XML dumps that are converted to the \emph{avro} format \cite{avro}, we ran a regular expression to extract the mentioned references in those revisions. As the scope of this study is limited to understanding the role of human editors in maintaining the reference quality of Wikipedia articles, we filtered out edits made by bots in the further analysis.

We built a dataset to examine the lifespan of deprecated and blacklisted domains. We froze the date to January 2022 and obtained the history of all references listed in the perennial sources list used until that point. The dataset consists of the following information for each occurrence of a reference: the page id, the timestamp when the reference was added, the timestamp when it was removed, the domain of the reference, the category of the domain, and the timestamp when the corresponding domain was classified as deprecated or blacklisted in the perennial source list, if applies. If a reference was added but not yet removed, the removal timestamp will remain blank.

\begin{itemize}
\item \textbf{Reference History} dataset consists of 4,203,467 occurrences of references that are still existing and that are removed.
\end{itemize}

\section{Methodology}

We quantify the reference quality of Wikipedia articles based on the following two criteria. First is the reference need (RN) that represents the proportion of sentences that need a citation, i.e., the information that is "likely to be challenged" \cite{verifiability} by readers. We release Citation Detective, a new tool that automatically identifies sentences needing citations given any revision. Second is the reference risk (RR) representing the proportion of sentences with unreliable citations. To compute this metric, we examine each revision and identify sources marked by the Wikipedia community as problematic in the perennial sources list~\cite{PerennialSource}. Figure~\ref{fig:idea} depicts the flow of our approach.

\subsection{Reference Need (RN)}

The RN metric operationalizes the reference quality of articles by computing the percentage of unreferenced sentences among those that need a citation. We describe the Citation Detective tool, which uses machine learning to compute this score for a given sentence. 

\subsubsection{Citation Detective}

Our new tool is based on previous research that was aimed at spotting sentences needing citations based on a machine-learning classifier~\cite{redi2019citation}. For example, the sentence ``It is estimated that throughout Florida, the storm-damaged 101,241 homes and destroyed approximately 63,000 others'' is marked as needing a citation by the classifier since it contains statistics~\cite{HA}.

We expanded on the idea of handling larger data and packaged it into an end-to-end inference pipeline. \textit{Citation Detective} is usable by the broader Wikipedia community and beyond. It currently monitors the English edition of Wikipedia and periodically releases a public dataset of unsourced statements on the encyclopedia. We release the code on github\footnote{For Citation Detective code refer to: \url{https://github.com/AikoChou/citationdetective}}. The tool proceeds as follows:
\begin{itemize}
   \item \textbf{Step-1 Retrieve the article content} based on the MediaWiki API\cite{WikiAPI}. Computation is scaled up with a pool of worker processes to parallelize the task.
 
   \item \textbf{Step-2 Segment and embed the article.} Every article is broken into sections and sentences. Sentences not needing citation, such as those under the "See also", "References", and "External links" sections, are discarded. Sentences and their section titles are split into words, where each word is transformed into its FastText embedding~\cite{joulin2016fasttext}.\footnote{FastText embeddings were chosen over the original GloVE embeddings~\cite{pennington2014glove} because of their ability to represent rare words. Moreover, FastText embeddings were available in multiple languages~\cite{fasttext}, which makes the Citation Detective system replicable for other language editions of Wikis. 
   Future studies can utilize newer language models like BERT and GPT2 for fine-tuning for the Wikipedia context.} 
   
   \item \textbf{Step-3 Infer Citation Need} For each sentence, the word embeddings and the section embeddings are forwarded into the Citation Need model~\cite{redi2019citation}. The output of this step is a score $\hat{y}$ in the range $[0,1]$: the higher the score, the more likely it is that a sentence needs a citation.
   
   \item \textbf{Step-4 Store results into a database.} For each sentence with a score higher than $\hat{y}=0.5$ (i.e., needing citation), we store the sentence text, the paragraph text that contains the corresponding sentence, the section title, the revision id of the article, and the predicted citation need score into an SQL database. The threshold is adjustable, but a value of 0.5 produces comparable outcomes to previous research on Wikipedia's reliability~\cite{lemmerich2019world}.
\end{itemize}

\subsubsection{Metric Definition for RN}

The Citation Detective tool classifies all sentences in a revision and labels each sentence with a binary Citation Need label $y$ according to the model output: $y=[\hat {y}]$, where $[\cdot]$ is the rounding function and $\hat{y}$ is the output of the Citation Need model. When $y=1$, the sentence needs a citation; when $y=0$, the sentence does not need one.

Next, we aggregate sentence-level Citation Need labels to compute each revision's RN score:
\begin{equation}
RN= 1 - \frac {1}{|P|}\sum _{i\in P}c_i,
\end{equation}
where $P$ is the set of sentences needing citations for a given article, i.e., having $y=1$; $c_i$ reflects the presence of a citation in the original text of the sentence $i$: $c=0$ if the sentence does not have an inline citation in the original text or $c=1$ if the sentence has an inline citation in the original text.

Once the RN scores of all revisions are computed for each revision, we can calculate the difference in RN scores ($dRN$) between revisions. The procedure for computing $dRN$ is as follows:
\begin{itemize}
   \item Each page in the dataset is sorted according to its revisions' timestamp.
   
   \item For each revision $r_t$ at time $t$ its difference on Reference Need is calculated as a difference between RN scores of $r_t$ and $r_{t-1}$.
   \begin{equation}
   dRN =RN_{r_t} - RN_{r_{t-1}},
   \end{equation}
   where $RN_{r_t}$ is RN score of ${r_t}$ and $RN_{r_{t-1}}$ is RN score of ${r_{t-1}}$
\end{itemize}
$dRN$ ranges between $[-1,1]$, with negative values meaning that reference need of an article improved after the revision (\emph{ie}., $RN$ score decreased) and positive values, vice versa.

\subsection{Reference Risk (RR)}

\subsubsection{Perennial sources list.} 

We focus on the precision of the two non-reliable categories of the perennial source list: deprecated and blacklisted. To motivate the choice of the Wikipedia community's internal labeling, we compare the perennial sources list with other external lists. We refer to UC Berkeley’s guide to fake news~\cite{berkeleylib} to identify well-established directories of unreliable sources. Three lists are used: (1) Zimdar’s `False, Misleading, Clickbait-y, and Satirical "News" Sources’ \cite{zimdars2016false}, (2) Daily Dots \cite{dailydots2016false}, which updates Zimdar’s list, and (3) Snopes~\cite{snopes2016fake}.
Coverage of these external lists and Wikipedia’s perennial sources was computed as the percentage of articles referencing at least one of the domains from each list. We computed the coverage before the perennial source list was created (i.e., before 2018) and in the \textit{Current} dataset to show the latest Wikipedia coverage.

\begin{table}[h!]
\caption{Coverage of Wikipedia by the perennial sources list and external lists. With a fewer number of domains, the perennial sources cover a greater percentage of articles.}
\vspace*{-3mm}
\centering
\label{tbl:coverage}
\resizebox{0.92\columnwidth}{!}{
\begin{tabular}{c|c|c|c}
\toprule
\textbf{List}                                                       & \textbf{\begin{tabular}[c]{@{}c@{}}No. of\\ domains\end{tabular}} & \textbf{\begin{tabular}[c]{@{}c@{}}Coverage\\ (before)\end{tabular}} & \textbf{\begin{tabular}[c]{@{}c@{}}Coverage\\ (after)\end{tabular}} \\ \midrule
Perennial (all) & 356                                                               & 17.4\%                                                               & 20.1\%                                                              \\
Perennial (risky) & 54                                                                & 0.95\%                                                               & 0.17\%                                                              \\
Zimdar's                                                            & 791                                                               & 0.65\%                                                               & 0.65\%                                                              \\
Daily dots                                                          & 175                                                               & \textless{}0.1\%                                                     & \textless{}0.001\%                                                  \\
Snopes                                                              & 25                                                                & \textless{}0.001\%                                                     & \textless{}0.001\%                                                  \\ \bottomrule
\end{tabular}
}
\end{table}

Table~\ref{tbl:coverage} shows each list's domains and coverage before and after the perennial sources list was created. There are fewer Wikipedia references on external lists than on lists from permanent sources. The coverage of persistently risky sources has dropped by a factor of five over time, which is a response from the Wikipedia community. On the other hand, external list coverage is almost unchanged. This motivated us to analyze the Wikipedia community-driven list.

\subsubsection{Metric Definition for RR.}
Using the perennial sources list, the article revision's $RR$ score is computed as the proportion of sentences containing unreliable references in that revision:
\begin{equation}
RR= \frac {x}{N},
\end{equation}
where $N$ is the total number of citations in a given revision; $x$ is the number of sentences pointing to an unreliable reference. Revisions not including any reference are omitted in this analysis. 

We compute the difference in the RR score ($dRR$) to monitor the impact of a given revision on the reference risk score. The negative $dRR$ value indicates an increase in an article's reliability (\emph{ie.}, $RR$ decreased) and the positive value, vice versa,
\begin{equation}
 dRR=RR_{r_t} - RR_{r_{t-1}},
\end{equation}
where $RR_{r_t}$ is the proportion of sentences with unreliable citations in ${r_t}$ and $RR_{r_{t-1}}$ is the proportion of sentences with unreliable citations in ${r_{t-1}}$.

\begin{figure*}[t!]
\centering
\captionsetup[subfigure]{labelformat=r-parens}
\begin{subfigure}[T]{0.63\textwidth}
       \includegraphics[width=0.48\linewidth]{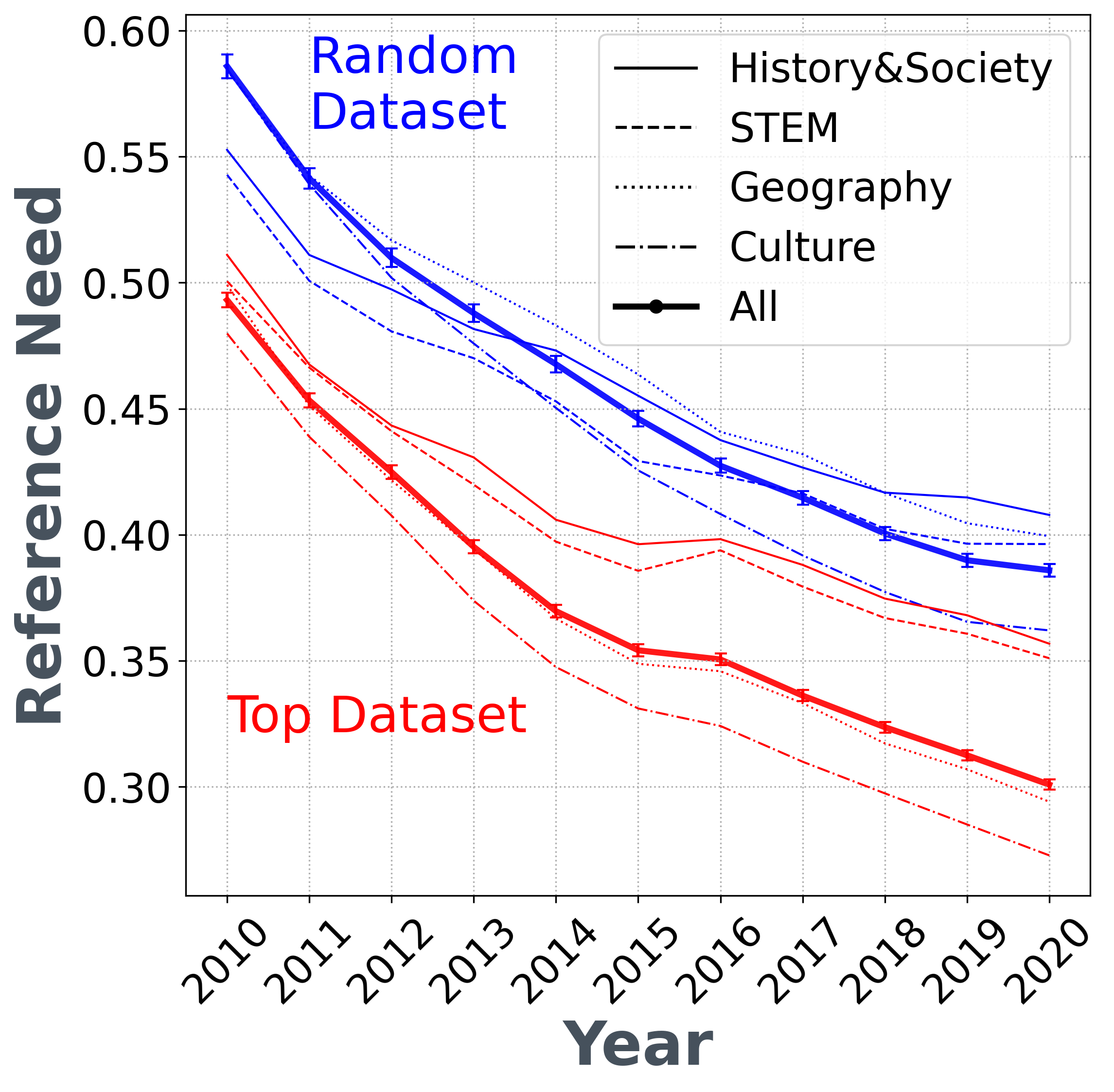}
       \includegraphics[width=0.49\linewidth]{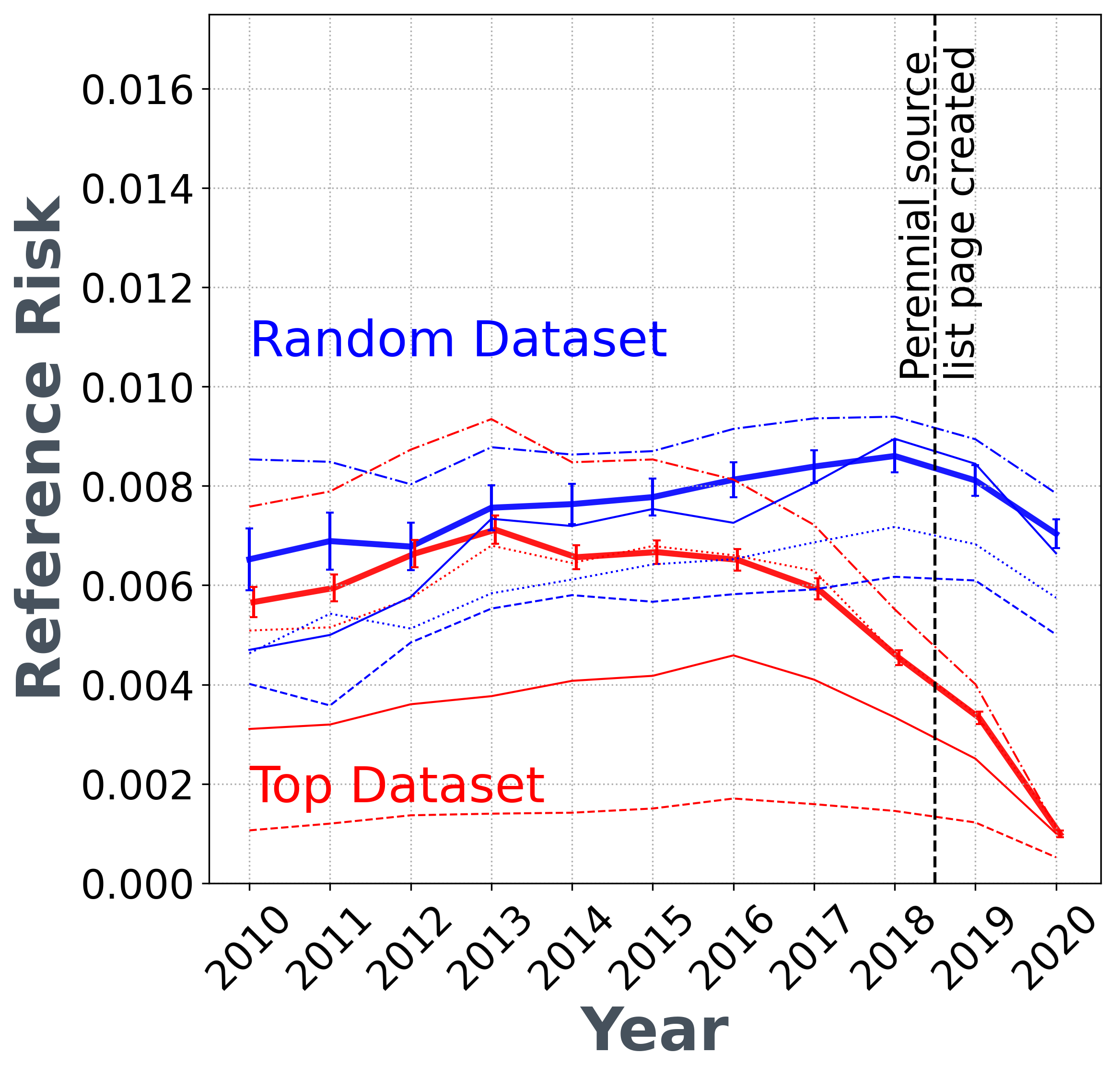}
      \caption{The evolution of reference quality in Wikipedia. Error bars represent the standard error. (Left) Reference need (RN) scores gradually decreased over the last decade, indicating an improved reference coverage of articles. The drop is nearly 20 percent point over the decade. (Right) Reference risk (RR) scores remain under 1\% and show a decreasing trend in recent years, suggesting a reduction in the use of risky references after the introduction of the perennial sources list in 2018.}
      \label{fig:evolution}
\end{subfigure}
\Description[Evolution of Reference Quality in Wikipedia]{Figure 2a consists of two line charts. The x-axis is a year ranging from 2010 to 2020. The y-axis in the left subfigure is the Reference Need score, and the right subfigure is the Reference Risk score. There are separate lines for the Top dataset (red) and the Random dataset (blue). The figure shows the evolution of RN and RR by topical category for each dataset, including History \& Society, STEM, Geography, and Culture. These are plotted in different line styles, and the scores, including All categories, are shown in bold. In the left subfigure showing RN evolution, all five lines for the Top dataset gradually decreased with an overall drop from 0.5 to 0.3, while all five lines for the Random dataset decreased from 0.6 to 0.4.
In the right subfigure showing RR evolution, for the Top dataset, the RR score fluctuated within the range of 0.004-0.007 before 2018 and then decreased, reaching 0.001 in 2020. For the Random dataset, the RR score gradually increased from 0.006 to 0.008 until 2018 and then decreased from 0.008 to 0.005. To indicate the creation of the perennial sources list page in 2018, there is a vertical dotted line between 2018 and 2019.
}
\hfill
\begin{subfigure}[T]{0.355\textwidth}
        \includegraphics[width=.99\linewidth]{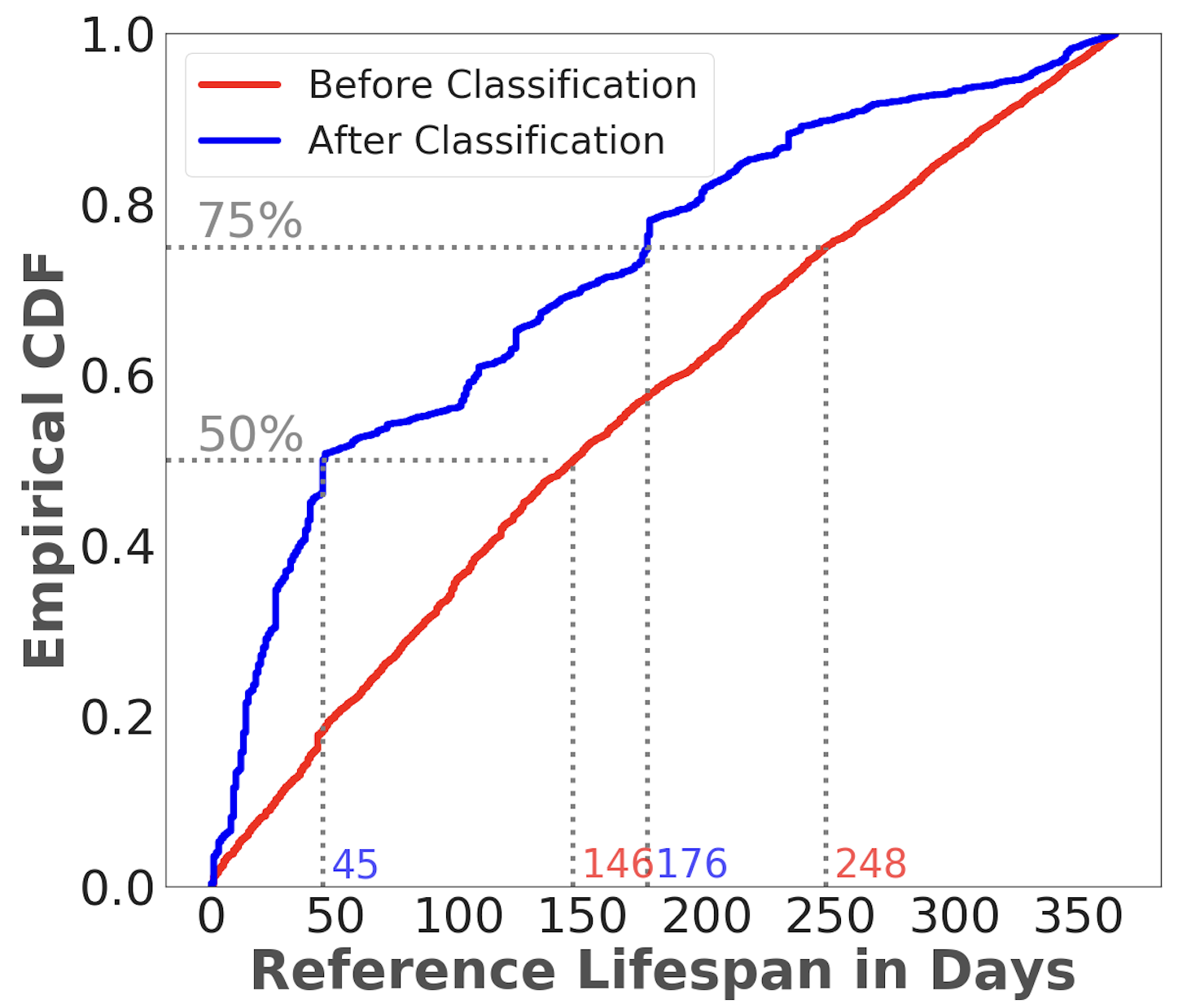}
        \Description[Lifespan of unreliable sources before and after their classification in the perennial sources list]{Line chart of empirical CDF for the lifespan of unreliable sources with the y-axis ranging between 0.0-1.0. The x-axis of the chart is the lifespan of references in days ranging from 0 to 350. There are two empirical CDF lines for Before Classification (red) and After Classification (blue). For red and blue lines, two dotted lines show the lifespan of reference in 50\% and 75\% population. Before classification, the lifespan of references in 50\% and 75\% populations is 146 days and 248 days, respectively. After classification, the lifespan in 50\% and 75\% populations is 45 days and 176 days, respectively.}
        \caption{The lifespan of unreliable sources a year before and after being added to the perennial sources list. Sources have a short lifespan on Wikipedia once marked as unreliable. \looseness=-1}
        \label{fig:perlist_lifespan}
\end{subfigure}
\end{figure*}

\subsection{Editor Expertise}

We will later propose a method for increasing the reliability of references by examining the role of participating editors. We use past editing history as a proxy for each editor's \textit{experience level} and divide editors into expert and novice categories based on the number of prior revisions.
\begin{itemize}
    \item \textbf{Experts} are editors who have contributed to a substantial number of revisions in the past, i.e., whose revision count is higher than the third quartile of the distribution of editors' prior revision counts.

    \item \textbf{Novices} are editors who have revised relatively few articles, i.e., whose revision count is below the first quartile in the distribution of editors' prior revision counts.
\end{itemize} 

We use a quasi-experimental setting by performing propensity score matching (PSM) and matching revisions of comparable editing environments. PSM uses logistic regression to assign propensity scores and is a statistical matching technique to estimate the effect of a treatment by accounting for the covariates~\cite{austin2011introduction}. We implemented with \textit{Pymatch}, a Python package tailored for propensity score matching. Detailed matching methods are described in Section 5.3. We plan to share the analysis code for PSM.

\section{Results}

We analyze the collected data to quantify the reference quality across time, topics, and editors.

\subsection{The Evolution of Reference Quality}

\subsubsection{Evolution of Reference Need}
Tracking the RN and RR scores allows us to examine how reference quality has evolved over the past decade. Shown in the left plot of Figure~\ref{fig:evolution} is the evolution of the reference need score. First, the average reference need score per article went down gradually over the last ten years, dropping by around 20\% in both \textit{Top} and \textit{Random} datasets. This demonstrates that a greater proportion of Wikipedia pages now include citations, or more than 60\% of citation-requiring sentences accompany a reference.

\subsubsection{Evolution of Reference Risk}
The evolution of the reference risk score is shown in the right plot of Figure~\ref{fig:evolution}. The risk score has remained below 1\% throughout the analyzed period. While the score only started to decrease in 2018 for the \textit{Random} dataset, the \textit{Top} dataset saw a decline starting in 2016. The decrease in the RR score coincided with the introduction of the perennial source list in 2018. This might suggest that the collaborative effort of Wikipedia editors enabled them to address newly registered non-authoritative sources, resulting in a decrease in the following years. We observe that the RR scores across the two datasets have increasingly diverged over the past few years.

\subsubsection{Reference Quality and Article Popularity}

Articles in the \textit{Top} dataset have a higher reference quality (i.e., lower RN and RR scores) than those in \textit{Random}. To confirm this observation for both indicators, we use the odds ratio to check if the pages with a higher view count have higher odds of being improved compared to a random sample. The odds ratio is a statistical measure of the association between two events. The events in our case are that (1) a page is top-viewed and (2) the page's reference quality has improved. By reference quality improvement, we mean how the page's RN and RR scores changed from the first to the last revision over the ten years under consideration. If the score decreases, we label the page as being improved; otherwise, if the score is either unchanged or increased, we regard the page as not being improved. The analysis shows that the odds ratio of top-viewed articles to be improved over randomly sampled articles is 9.61 for RR and 2.65 for RN. This indicates an association between the number of page views and the improvement of the reference quality of the page.

\subsubsection{Lifespan of Risky Sources}

To explore the role of community-driven work in the evolution of reference quality, we examine whether classifying sources in the perennial source list as "deprecated" or "blacklisted" motivates editors to remove existing risky references. We calculate the lifespan of risky references as the time elapsed between their addition and removal using the \textit{Reference History} dataset. We analyze the lifespan of references within a year before and after their classification in the perennial sources list, as the list was established in 2018.

Figure~\ref{fig:perlist_lifespan} shows the median lifespan (or the number of days a reference survives before being removed by future edits) of risky references decreased by more than threefold once they were added to the perennial list by editors. Additionally, the lifespan of risky references at the 75th percentile decreases by approximately two months. These results indicate that labeling of perennial sources encouraged editors to remove unreliable references if they were labeled undesirable quickly. There was no definite consensus among deprecated sources regarding the domains of "Daily Mail" and "The Sun." Because their status was the subject of multiple discussions, they were excluded from our main analysis. Results including these domains are shown in Figure 5 in the Appendix.

\subsection{Reference Quality Across Topics}

\begin{figure}[t!]
    \centering 
    \includegraphics[width=\linewidth]{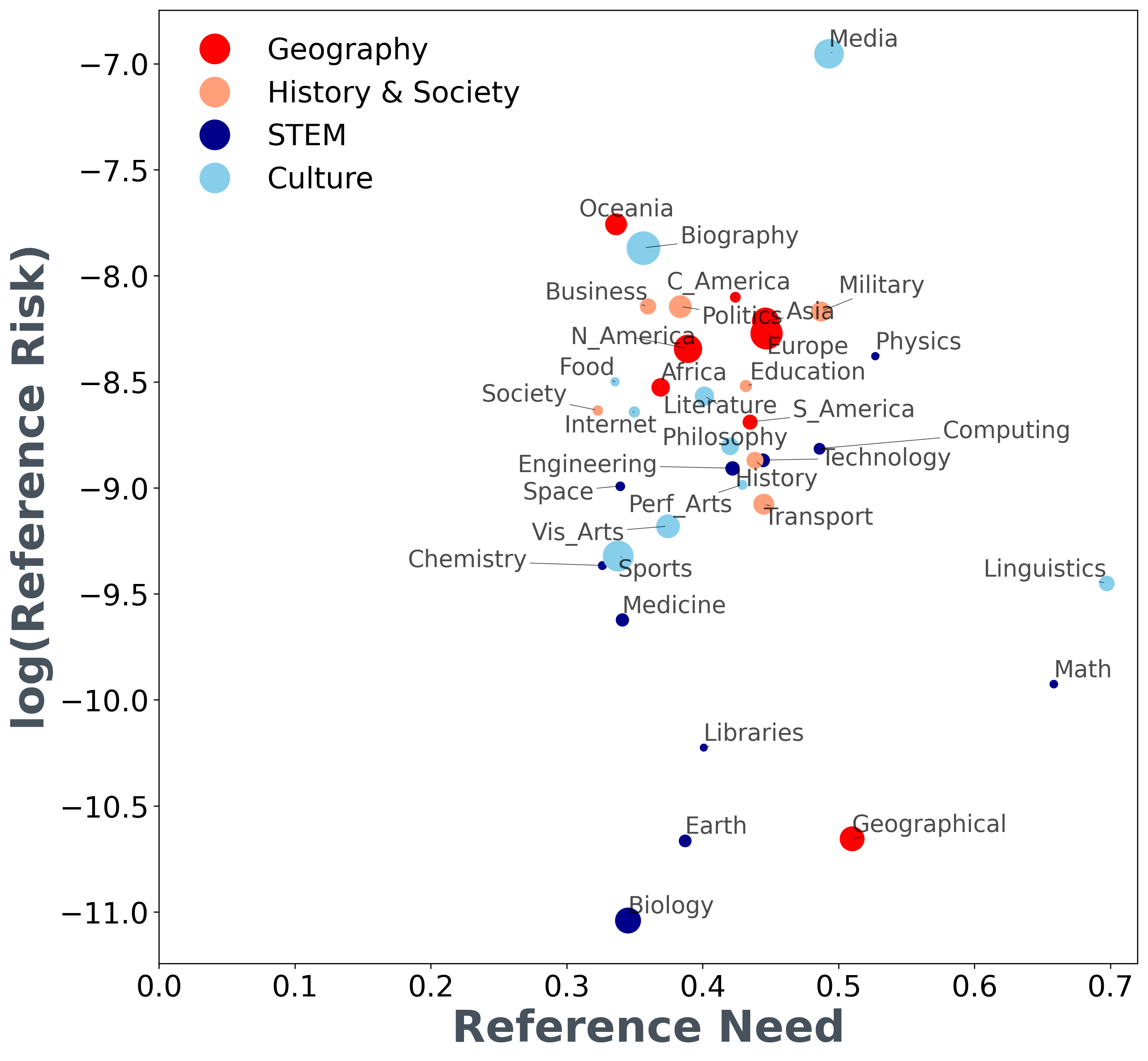}
    \Description[Distribution of topical categories by Reference Risk and Reference Need scores]{A scatter plot shows a relative position of topical categories by RN and RR scores. The x-axis of the chart is RN  (ranges from 0.0 to 0.7), and the y-axis is the logarithm of RR (ranges from -11.0 to -7). The scatter plot is shown in four colors, each representing a different topical cluster: red for Geography, coral for History \& Society, blue for STEM, and sky blue for Culture. The radius of each circle differs depending on the number of articles on that topic.}
     \vspace*{-3mm}
    \caption{The relative position of Wikipedia page clusters in the \textit{Current} dataset by the RR and RN score plane. The radius of each circle represents the number of articles in that cluster. Clusters in the bottom left corner generally have higher reference quality in terms of the RN and RR scores.} 
    \label{fig:topics}
\end{figure}

Compared to earlier studies that provided insights on claim verification in specific domains~\cite{Wadden2020FactOF, Wadden2022SciFactOpenTO}, we can observe reliability across a wider set of topics on Wikipedia. The reference quality scores differ by the topic of the articles. Figure~\ref{fig:topics} shows the distribution of reference quality for different topic clusters in the \textit{Current} dataset, where colors represent the higher level category.
Risky citations are more prevalent in articles related to Culture, such as media and biographies. Other topical clusters, such as military, politics, and Asia, are similarly subject to lower RN and RR scores. In contrast, articles related to STEM domains such as biology, earth, and libraries tend to contain fewer risky references.

Scientific domains like mathematics, physics, and linguistics tend to have higher RN values. This means that Wikipedia pages about these topics are more likely to have citations that need to be included. On the other hand, topics about society have the lowest RN values, indicating that they are well-referenced. Cultural topics tend to have lower RN scores and higher RR scores, while STEM-related articles have higher RN scores and lower RR scores. This aligns with the evolution of reference quality by topical category in Figure~\ref{fig:evolution}. The RN score reflects the machine-calculated indicator of reference need and not necessarily the human perception of reference need. We conjecture that Wikipedia readers may be less concerned about missing references for specific topics.

None of the top 10 articles that received the most page views in each topical category contained unreliable sources. The average RN scores for top-viewed articles were substantially lower. These findings indicate that top-viewed pages have a higher reference quality, likely because top-viewed content is regularly visited by editors, increasing their chances of further edits.

\subsection{Reference Quality Across Editors}

\begin{table*}[t!]
\caption{The average difference in reference need and risk scores. (1) Expertise (i.e., expert and novice editors): contents edited by experienced users result in reduced reference risk and need scores. (2) Interaction (i.e., exposed and unexposed novice editors): contents edited by novice users exposed to experts have a lower increase in reference risk and need scores. 
}
\label{tbl:psm_table}
\centering
\resizebox{0.9\textwidth}{!}{%
\begin{tabular}{cc|cccccc|ccccll}
\toprule
\multirow{3}{*}{\textbf{Data}} &
  \multirow{3}{*}{\textbf{Metric}} &
  \multicolumn{6}{c|}{\textbf{Expertise}} &
  \multicolumn{6}{c}{\textbf{Interaction}} \\
  \cline{3-14}
 &
   &
  \multicolumn{2}{c}{\textbf{Experienced}} &
  \multicolumn{2}{c}{\textbf{Novice}} &
  \multirow{2}{*}{\textbf{t}} &
  \multirow{2}{*}{\textbf{p-value}} &
  \multicolumn{2}{c}{\textbf{Exposed}} &
  \multicolumn{2}{c}{\textbf{Unexposed}} &
  \multicolumn{1}{c}{\multirow{2}{*}{\textbf{t}}} &
  \multicolumn{1}{c}{\multirow{2}{*}{\textbf{p-value}}} \\ \cline{3-4} \cline{5-6} \cline{9-12}
 &
   &
  Mean &
  SD &
  Mean &
  SD &
   &
   &
  Mean &
  SD &
  Mean &
  SD &
  \multicolumn{1}{c}{} &
  \multicolumn{1}{c}{} \\ \midrule
\multirow{2}{*}{\textbf{Top}} &
  \textbf{$dRN$} &
   -0.007 &
   0.089 & 
   0.002 &
   0.084 &
   -25.3 &
   <0.001 &
  0.002 & 
  0.083 &
  0.004 &
  0.083
   & -5.4 
   & <0.001
   \\
 &
  \textbf{$dRR$} &
  -0.001 &
   0.012 &
  -0.000 &
   0.012 &
   -6.07&
   <0.001&
  0.000 & 0.010
   & 0.000 & 0.011
   & 2.3
   & <0.05
   \\ \cline{2-14} 
\multirow{2}{*}{\textbf{Random}} &
  \textbf{$dRN$} &
  -0.013 &
   0.190 &
   0.016 &
   0.183&
   -16.9&
   <0.001&
  0.012 & 0.17
   & 0.025 &  0.19
   & -6.2
   & <0.001
   \\
 &
  \textbf{$dRR$} &
  -0.001 &
   0.074 &
  -0.003 &
   0.0 &
   2.68 &
   <0.01
   &  0.000 & 0.064
   &  0.004 & 0.073
   & -5.1
   & <0.001
   \\ \bottomrule
\end{tabular}%
}
\end{table*}
 
To understand what impacts reference quality, we now examine how the editor's prior experiences and the act of co-editing impact the RR and RN metrics.

\subsubsection{Editor Experience and Reference Quality}

To match the edits of novice and expert editors, we used three covariates that directly affect the quality of a revision: (i) the size of the revision in bytes, (ii) the reference quality scores of the article prior to the revision, and (iii) the content topic. Covariates are specified to compute propensity scores via logistic regression. With the propensity scores, we employed one-to-one revision matching to create the control and test groups. Edits that have no effect on the reference quality of the article were not considered for matching. To check the resultant groups for balance, we used Standardized Mean Difference (SMD) for numerical covariates and a Chi-square independence test for categorical covariates. We report the results in the Appendix.

After matching revisions, we compared the average change in reference quality to the editor's experience in test and control groups. T-test was used to measure the effect of expertise on reference quality, which suggested substantial differences in reference quality between the two groups, as summarized in Table~\ref{tbl:psm_table} (see the expertise tab). Experienced editors show a negative change in RN and RR scores, indicating that the edited content is of higher quality. In contrast, the score changes for novice editors are positive and close to zero, except for the RN score in the \textit{Random} dataset. This means that the content they edited may need to be changed again later to include the needed references, or that the references themselves need to be updated.

\subsubsection{Co-editing and Reference Quality}

Can the data tell us the effect of editing an article with another editor on the same day (which we call ``co-editing'') on future edits? In particular, we are interested in whether novice editors see any benefit from co-editing with a more experienced editor. Along with the three covariates used earlier, we now employ another covariate to match the edits of novice users based on co-editing experience with experts: (iv) revision count of the editor prior to the given revision. All other settings for the PSM analysis are the same as before.

Novice users were separated into two groups based on their co-editing experience. The first group consists of `exposed' users who have at least once co-edited an article with an experienced editor on the same day, which is the control group for PSM. Users in the second group are `unexposed' and have never co-edited with experienced editors on the same article, which is the test group. Editing the same article on the same day is a reasonable indicator of co-editing, as editors are notified if their edits are reverted~\cite{revert}. After matching the two groups' revisions, we examined how the `exposed' and `unexposed' novice editors differ in the quality of their edited content. T-test was conducted to verify the significance of the distribution difference. Table~\ref{tbl:psm_table} indicates that revisions made by novice users who were exposed to experienced editors are less likely to require references or contain risky references. 

\subsubsection{Alternative Matching}
Other matching methods, such as Mahalanobis distance matching (MDM), may be considered for the robustness check. With the use of MDM, we obtained consistent results regarding the editor experience and co-editing experiments on reference quality, as the findings above indicate. These results are given in the Appendix.

\section{Discussion and Conclusions}

So far, we have conducted large-scale assessments of the reference quality of Wikipedia articles
by defining two quality factors: Reference Need (RN) and Reference Risk (RR). For the former, we also provide a tool to monitor the need for citations on a scale that editors and researchers can use. For the latter, we designed a methodology based on a community-created reliability index, discovering relevant insights about the usage and growth of non-reliable sources in the English edition of Wikipedia. \smallskip 

\subsection{A Decade-Long History of Improving Reference Quality on Wikipedia}

The RN index gradually decreased over the past decade, indicating that more articles now accompany references. This trend results from an increasing volume of community initiatives aimed at improving citation coverage, including the exceptional work done by editors and the success of tools such as Citation Hunt \cite{citationhunt} that aims at resolving the $[\textit{citation needed}]$ tags, as well as initiatives such as 1Lib1Ref \cite{1Lib1Ref} that invites librarians to add citations to Wikipedia pages. These efforts improve Wikipedia itself and, in return, ensure a higher quality encyclopedia for humans and machines. According to~\cite{piccardi2020quantifying}, only 0.3\% of page visitors click on citations. This means that only rarely are sentences verified by readers, making the role of editors critical to ensuring knowledge integrity and verifiability.

Our results may be considered a lower bound of the reference risk value because the Perennial Sources list only covers a small fraction of potentially unreliable sources. Unfortunately, using external reliability indexes and fact-checking systems such as Full Fact \cite{fullfact} is difficult in the Wikipedia context, given that existing lists are country-specific or not generic enough to cover the diversity of topics and sources. Creating a global index of source reliability would improve this estimate, support targeted interventions in specific content areas, and expose potential disinformation attacks from malicious users. Together with other efforts to build trust around the world, our scientific community could support such a global effort to improve and keep an eye on the quality of Wikipedia's sources that directly affect the services people use.
Systems that help automatically flag the presence of newly added unreliable sources could help editors monitor reference quality, and this paper provides a foundational methodology to build such support tools.

\subsection{Reference Quality Varies by Topic}
Focusing on the distribution of reference quality over topics, scientific domains like Physics and Mathematics have high reference reliability but high reference need. Similar to other scientific topics (e.g., Medicine), reference risk on these subjects is low due to the high reliability of scientific journals cited. One reason Mathematics and Physics are marked as poorly sourced may be that they do not report inline citations for all claims detected as needing citations, likely because these disciplines verify text claims using formulas, equations, and plots accompanying the text.

Biographies are another important topic to observe, as they make up around 30\% of Wikipedia articles and follow special referencing rules and guidelines. For example, the Wikimedia Foundation's resolution on Biographies of Living People~\cite{BiograpyWiki}, one of the rare cases that interfere with the projects' content, encourages editors to make sure that this class of articles is ``neutrally-written, accurate, and well-sourced.'' Our analysis shows that, while they are generally well-covered in terms of references, Biographies are one of the topics with the highest reference risk. We plan to investigate the specific distribution of reference risk across demographic traits to expose potential biases in reference quality and provide tools for editors to identify unreliable sources in biographies.

\subsection{Co-editing Experience Can Help Improve Reference Quality}

Our quasi-experimental analysis based on PSM indicated a statistically significant difference in the reference quality of revisions made by experienced and novice editors. Following previous research~\cite{holtz2018effects,10.1145/1929916.1929920}, our data show that experienced editors are more likely to improve the reference quality of Wikipedia articles by supplementing edits with citations as well as removing and avoiding potentially risky references. This emphasizes the importance of initiatives for editor retention, such as the Structured Task program~\cite{StructuredTasks}, MoodBar~\cite{ciampaglia2015moodbar} or the Wikiproject Editor Retention~\cite{EditorRetention}. In the future, researchers can look at how long-term editor retention has changed over time.
The PSM analysis further suggests that novice editors might benefit from having edited the same articles with more experienced editors on the same day; those who co-edited an article later produced revisions with higher reference quality. This finding aligns with previous work that suggests the importance of creating healthy interactions between experienced and novice editors~\cite{morgan2013tea,halfaker2011don} and the crucial impact of collaboration among different types of editors on article quality~\cite{liu2011does}.

\subsection{Other Web Services Can Adopt the Participatory Verification Culture}

The collaborative nature of the encyclopedia is essential for ensuring high content reliability, as it enables editors to be more productive and improve the quality of the content. The picture is much different on other platforms of user-generated content, such as social media, that are designed to maximize engagement and attract public attention. Unlike Wikipedia, web content recommended or discussed by multiple individuals on social media is not necessarily more reliable, as shown in increasing cases of fake news~\cite{ross2018fake,naeem2021exploration}. In contrast, co-edited articles have become more reliable over time, as has the overall reliability. Other Web services may consider making user longevity and contributions to the platform more explicit, as in Wikipedia.

\subsection{Limitations and Future Work}

This research can be improved in numerous ways. One is the scope of the data. The current study focused on the English edition of Wikipedia without providing insights into the other 300+ language editions. We would like to extend this work and support more languages with the \textit{Citation Detective} tool. 
Future work could adapt the Citation Need model and tools and repeat the analysis for other Wikipedia editions. Such analyses can uncover patterns around how different language communities monitor and improve reference quality and provide insights that can help preserve the global integrity of Wikipedia knowledge.

Another limitation is the scope of the metric. Although the suggested metrics represent potentially harmful sources and uncited claims in an article, we cannot assure whether the referenced source supports the cited claim and to what extent. For example, the RN index does not capture the strength of the claim, nor can we detect whether a single source supports multiple claims. Similarly, the RR index does not capture whether the cited source actually contains evidence to support the respective claim. Future studies can incorporate claim verification as a measure of the quality metric.
The coverage of the perennial sources list for the RR index can be extended as well. While the list has been carefully compiled based on extensive community discussion, the number of sources included is small compared to the total number of sources used in Wikipedia. Developing a machine-supported list may help the community assess more sources.

For the correlation analysis, we examined trends over time, topical categories, and participating editors. We plan to examine a broader range of factors to identify better opportunities to improve Wikipedia's reference quality. In the future, researchers could look into more subtle differences between editors and add more clustering dimensions beyond the number of edits. For example, one may use a broader range of matching techniques to expand on the PSM and MDM methods.

We also highlight the need for creating new tools to assist the daily work of Wikipedia editors, facilitating their ability to find, assess, and fix references. We hope our work inspires researchers and developers to build new and more efficient tools to test the reliability of Wikipedia and other Web services.


\begin{acks}
We thank the track chairs and reviewers for their invaluable guidance and feedback.
This research was supported by the National Research Foundation (RS-2022-00165347) and Institute for Basic Science (IBS-R029-C2) in Korea. 
\end{acks}

\bibliographystyle{ACM-Reference-Format}
\typeout{}
\balance
\bibliography{main}

\appendix
 \clearpage
\section{Wikipedia page layout}
\begin{figure}[ht!]
\centering
  \includegraphics[width=\linewidth]{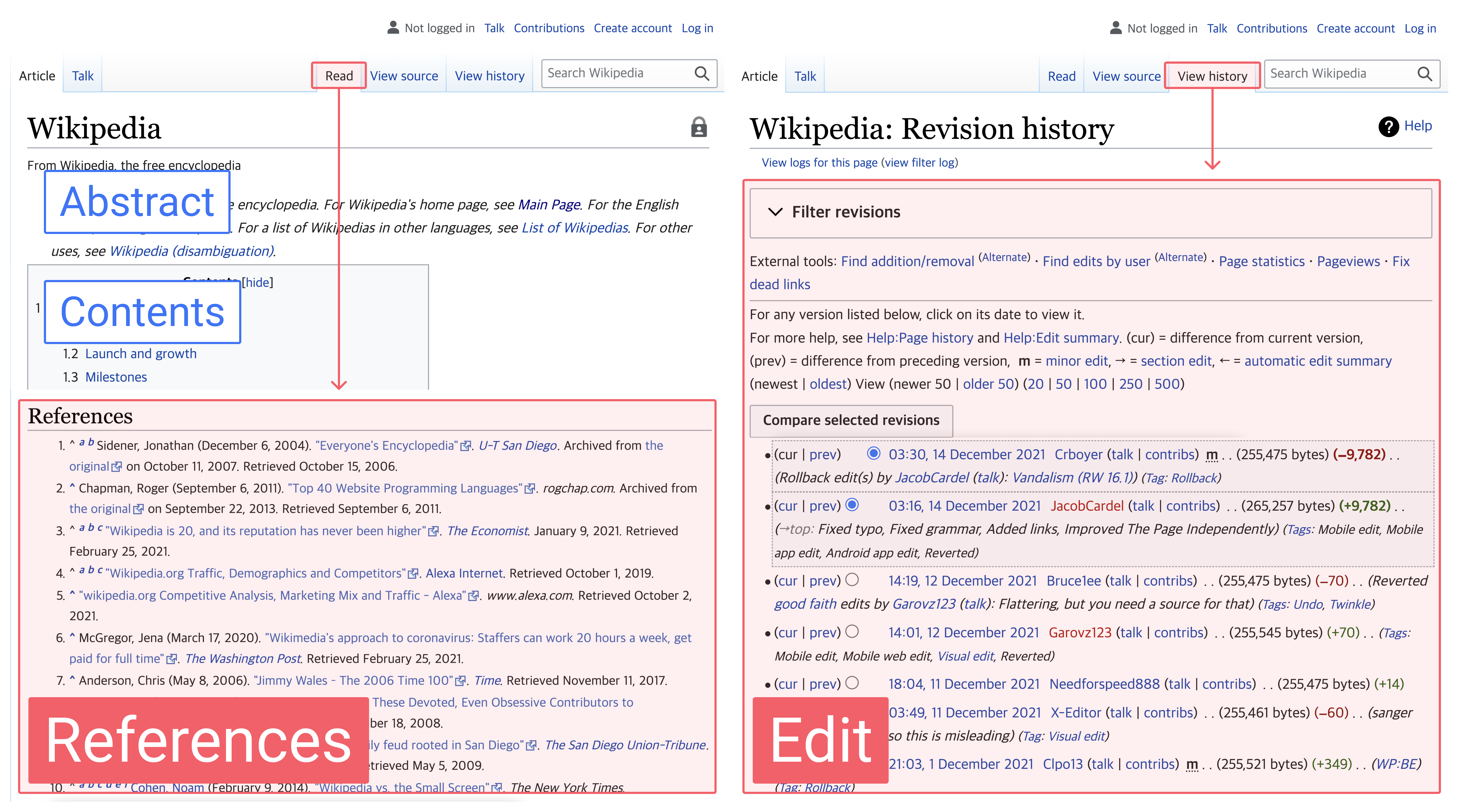}
  \Description[Layout of a Wikipedia page]{The screenshot of the 'Wikipedia' page taken from Wikipedia. There are two screenshots. On the left, the 'Read' tab is shown, and on the right, the 'View history' tab is shown. In the 'Read' tab, the top part is divided into 'abstract' and 'contents' by blue boxes, and the 'references' section is highlighted by a red box. In the 'View history' tab, the edit history of the page is shown and is highlighted by another red box.}
\caption{The layout of a Wikipedia page. The 'view history' tab lists all previous versions of the page with information on the editor, the timestamp of the edit, the change made by the revision in bytes, and the editor's comment.}
\label{fig:wiki_example}
\end{figure}

\section{Further results on the lifespan of Unreliable References}
\begin{figure}[h]
\centering
  \includegraphics[width=0.8\linewidth]{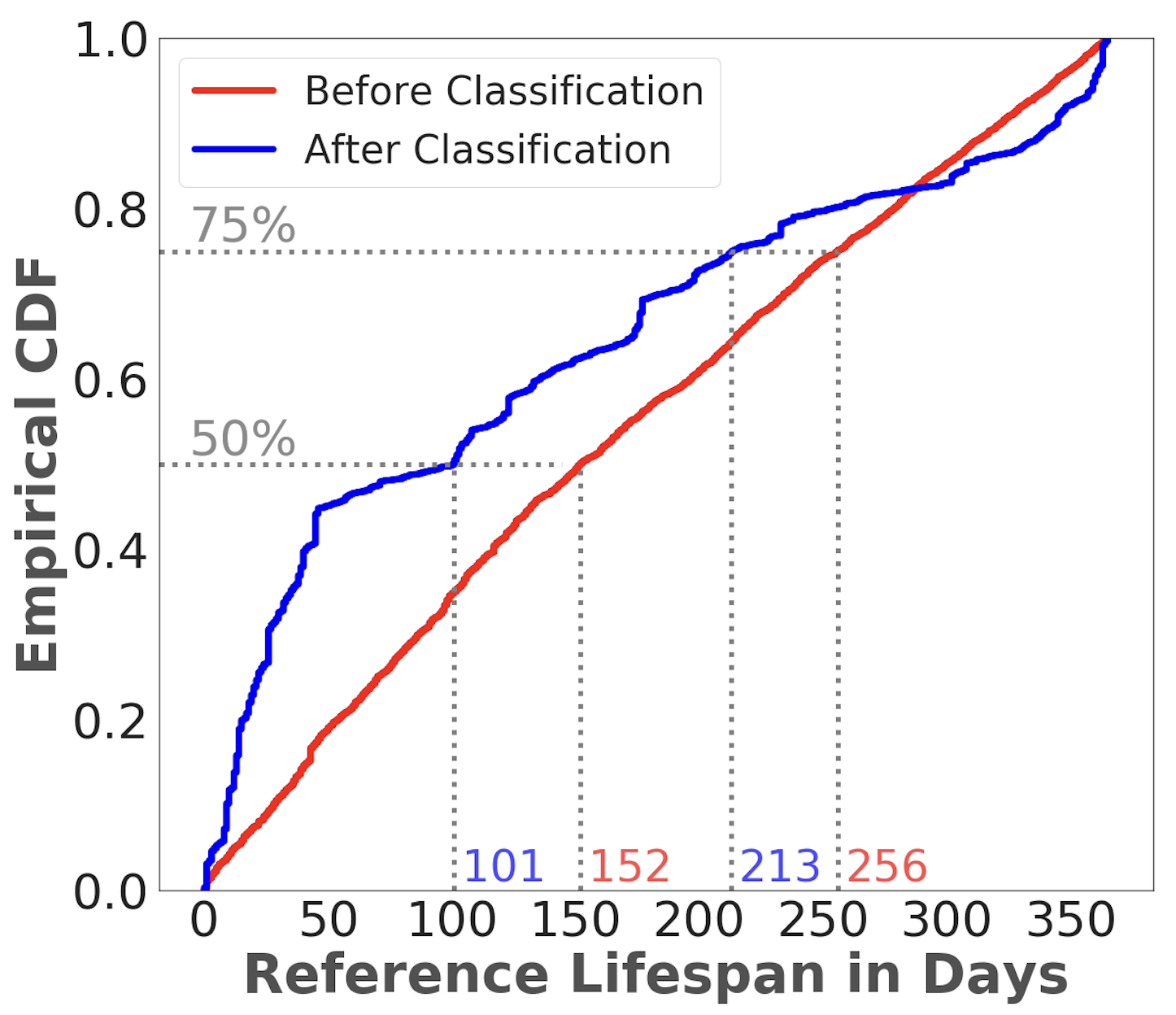}
  \Description[Lifespan of unreliable sources before and after their classification in the perennial sources list before filtering out 'The Sun' and 'Daily Mail']{Line chart of empirical CDF for the lifespan of unreliable sources with the y-axis ranging between 0.0-1.0. The x-axis of the chart is the lifespan of references in days ranging from 0 to 350. There are two empirical CDF lines for Before Classification (red) and After Classification (blue). For red and blue lines, two dotted lines show the lifespan of reference in 50\% and 75\% population. Before classification, the lifespan of references in 50\% and 75\% populations is 152 days and 256 days, respectively. After classification, the lifespan in 50\% and 75\% populations is 101 days and 213 days, respectively.}
\caption{The empirical distribution of the lifespan of unreliable sources. Two long-controversial sources, such as "The Daily Mail" and "The Sun", are mainly distributed in long-standing lifespan}
\label{fig:distributionlifespan_2}
\end{figure}

\section{Categories of the Perennial Sources List}

\begin{itemize}
    \item \textbf{Generally Reliable} (Not risky) - generally reliable in its areas of expertise. Example: abcnews.go.com
    \item \textbf{Non-Consensus} (Not risky) - marginally reliable (i.e. neither generally reliable nor generally unreliable) and may be usable depending on the context. Example: askmen.com
    \item \textbf{Generally Unreliable} (Not risky) - editors show a consensus that the source is questionable in most cases. Example: abcnews.go.com
    \item \textbf{Deprecated} (Risky) - considered generally unreliable, and use of the source is generally prohibited. Example: dailycaller.com 
    \item \textbf{Blacklisted} (Risky) - due to persistent abuse, usually in the form of external link spamming, the source is on the spam blacklist or the Wikimedia global spam blacklist. Example: bestgore.com
\end{itemize}

\section{PSM Balance}

To evaluate the data balance of pairs matched with PSM, Standardized Mean Difference (SMD) was used for the numerical covariates (previous score, revision size, and editor revision count), 
and a chi-square test was used for the categorical covariates (article topic). SMD is the common way to evaluate imbalance in covariate distribution, and its recommended threshold for declaring imbalance is 0.1~\cite{zhang2019balance}. Results are shown in Table \ref{tab:PSM-stats}. Reduced SMD values and the p-values >0.05 after matching indicate balanced data following Pymatch documentation.

\begin{table}[H]
    \caption{Assessment of data balance across covariates after performing propensity score matching.}
    \begin{subtable}[h]{\columnwidth}
    \centering
    \caption{Expertise}
    \label{tab:matching-stats-exp}
    \resizebox{\columnwidth}{!}{%
    \begin{tabular}{cc|cccccc}
    \hline
    \multirow{3}{*}{\textbf{Data}} & \multirow{3}{*}{\textbf{Metric}} & \multirow{3}{*}{\textbf{\# of Pairs}} & \multicolumn{4}{c}{\textbf{Std Mean Difference}} & \textbf{\begin{tabular}[c]{@{}c@{}}Chi-Squre\\ (p-value)\end{tabular}} \\ \cline{4-8} 
     &  &  & \multicolumn{2}{c}{\textbf{Prev. Score}} & \multicolumn{2}{c}{\textbf{Rev. Size}} & \multirow{2}{*}{\textbf{Topic}} \\
     &  &  & Before & After & Before & After &  \\ \hline
    \multirow{2}{*}{\textbf{Top}} & \textbf{dRN} & 184007 & 0.004 & -0.069 & 0.100 & 0.033 & 0.00 \\
     & \textbf{dRR} & 36672 & 0.010 & -0.007 & 0.136 & 0.030 & 0.00 \\ \hline
    \multirow{2}{*}{\textbf{Random}} & \textbf{dRN} & 27189 & -0.109 & -0.018 & -0.030 & -0.018 & 0.36 \\
     & \textbf{dRR} & 34773 & -0.080 & 0.028 & -0.037 & -0.011 & 0.41 \\ \hline
    \end{tabular}%
    }
    \end{subtable}
    \hfill
    \begin{subtable}[h]{\columnwidth}
    \centering
    \caption{Interaction}
    \label{tab:matching-stats-int}
    \resizebox{\columnwidth}{!}{%
    \begin{tabular}{cc|cccccccc}
    \hline
    \multirow{3}{*}{\textbf{Data}} & \multirow{3}{*}{\textbf{Metric}} & \multirow{3}{*}{\textbf{\# of Pairs}} & \multicolumn{6}{c}{\textbf{Std Mean Difference}} & \textbf{\begin{tabular}[c]{@{}c@{}}Chi-Square\\ (p-value)\end{tabular}} \\ \cline{4-10} 
     &  &  & \multicolumn{2}{c}{\textbf{Prev. Score}} & \multicolumn{2}{c}{\textbf{Rev. Size}} & \multicolumn{2}{c}{\textbf{User Rev. Count}} & \multirow{2}{*}{\textbf{Topic}} \\
     &  &  & Before & After & Before & After & Before & After &  \\ \hline
    \multirow{2}{*}{\textbf{Top}} & \textbf{dRN} & 113787 & -0.233 & -0.020 & 0.021 & 0.005 & 0.630 & -0.010 & 0.00 \\
     & \textbf{dRR} & 37823 & -0.044 & 0.022 & 0.053 & -0.000 & 0.654 & -0.001 & 0.00 \\ \hline
    \multirow{2}{*}{\textbf{Random}} & \textbf{dRN} & 15365 & -0.120 & -0.016 & 0.010 & 0.010 & 0.305 & 0.017 & 0.30 \\
     & \textbf{dRR} & 16560 & -0.086 & -0.006 & 0.020 & -0.005 & 0.345 & -0.017 & 0.09 \\ \hline
    \end{tabular}%
    }
    \end{subtable}
\label{tab:PSM-stats}
\end{table}

\section{Alternative matching}

\begin{table}[H]
    \caption{The average difference in reference need and risk scores using Mahalanobis distance matching.}
    \begin{subtable}[h]{\columnwidth}
    \centering
    \caption{By user expertise}
    \label{tbl:MDM-exp}
    \resizebox{\columnwidth}{!}{%
    \begin{tabular}{@{}cccccccc@{}}
    \toprule
    \multirow{2}{*}{\textbf{Data}} & \multirow{2}{*}{\textbf{Metric}} & \multirow{2}{*}{\textbf{\# of Pairs}} & \multicolumn{2}{c}{\textbf{Mean}} & \multirow{2}{*}{\textbf{\begin{tabular}[c]{@{}c@{}}p-value\\ (T-test)\end{tabular}}} & \multicolumn{2}{c}{\textbf{Std Mean Difference}} \\ \cmidrule(lr){4-5} \cmidrule(l){7-8} 
     &  &  & \textbf{Exp.} & \textbf{Nov.} &  & \textbf{Prev. Score} & \textbf{Rev. Size} \\ \midrule
    \multirow{2}{*}{\textbf{Top}} & \textbf{dRN} & 184008 & -0.010 & -0.008 & \textless 0.001 & 0.0038 & 0.0086 \\
     & \textbf{dRR} & 36672 & -5.2e-4 & 6.7e-5 & \textless 0.001 & 0.0014 & 0.0826 \\ \midrule
    \multirow{2}{*}{\textbf{Random}} & \textbf{dRN} & 27191 & -0.013 & 0.014 & \textless 0.001 & -0.0297 & -0.0059 \\
     & \textbf{dRR} & 36502 & -0.001 & 0.001 & \textless 0.01 & -0.0768 & -0.0194 \\ \bottomrule
    \end{tabular}%
    }
    \end{subtable}
\hfill
    \begin{subtable}[H]{\columnwidth}
    \centering
    \caption{By user interaction with experienced editors}
    \label{tbl:MDM-int}
    \resizebox{\columnwidth}{!}{%
    \begin{tabular}{@{}ccccccccc@{}}
    \toprule
    \multirow{2}{*}{\textbf{Data}} & \multirow{2}{*}{\textbf{Metric}} & \multirow{2}{*}{\textbf{\# of Pairs}} & \multicolumn{2}{c}{\textbf{Mean}} & \multirow{2}{*}{\textbf{\begin{tabular}[c]{@{}c@{}}p-value\\ (T-test)\end{tabular}}} & \multicolumn{3}{c}{\textbf{Std Mean Difference}} \\ \cmidrule(lr){4-5} \cmidrule(l){7-9} 
     &  &  & \textbf{Co-edit} & \textbf{\begin{tabular}[c]{@{}c@{}}Non\\ Co-edit\end{tabular}} &  & \textbf{Prev. Score} & \textbf{\begin{tabular}[c]{@{}c@{}}User\\ Rev. Count\end{tabular}} & \textbf{Rev. Size} \\ \midrule
    \multirow{2}{*}{\textbf{Top}} & \textbf{dRN} & 289837 & 0.005 & 0.008 & \textless 0.001 & -0.0042 & -0.0026 & -0.0096 \\
     & \textbf{dRR} & 47661 & 3.7e-4 & 2.5e-4 & 0.11 & 0.0062 & -0.0005 & -0.0197 \\ \midrule
    \multirow{2}{*}{\textbf{Random}} & \textbf{dRN} & 27191 & 0.011 & 0.016 & \textless 0.001 & -0.0317 & 0.0283 & 0.0091 \\
     & \textbf{dRR} & 10843 & 3.5e-4 & 0.002 & \textless 0.01 & 0.0129 & 0.0194 & 0.0006 \\ \bottomrule
    \end{tabular}%
    }
    \end{subtable}
    \label{tab:MDM}
\end{table}

\end{document}